\documentclass[12pt]{article}
\newcommand{\as}{{\ifmmode \alpha_{\rm s} \else $\alpha_{\rm s}$ \fi}}
\newcommand{\ep}{\epsilon}
\newcommand{\be}{\begin{equation}}
\newcommand{\ee}{\end{equation}}
\newcommand{\bea}{\begin{eqnarray}}
\newcommand{\eea}{\end{eqnarray}}
\newcommand{\beas}{\begin{eqnarray*}}
\newcommand{\eeas}{\end{eqnarray*}}
\newcommand{\lspage}[1]   
{ 
\rotatebox{90}{ \begin{minipage}{\textheight} #1 \end{minipage} } 
  }
\def\np#1#2#3{
        {\it Nucl. Phys. }{\bf #1} (19#3) #2}
\def\pl#1#2#3{
        {\it Phys. Lett. }{\bf #1} (19#3) #2} 
\def\pr#1#2#3{
        {\it Phys. Rev. }{\bf #1} (19#3) #2} 
\def\prl#1#2#3{
        {\it Phys. Rev. Lett. }{\bf #1} (19#3) #2}

\usepackage{graphics}
\usepackage{axodraw}

\begin{document}
\begin{titlepage}
\begin{flushright}
GUELPH/99-296 \\
November 1999
\end{flushright}
\vspace{1.5cm}
\begin{center}
\Large
{\bf Split dimensional regularization for the Coulomb gauge at two loops}

\vspace{1.2cm}
\large
G.~Heinrich, G.~Leibbrandt \\

\vspace{1.5cm}
\normalsize
{\it Department of Mathematics and Statistics, University of Guelph,\\ 
Guelph, Ontario N1G 2W1\\
 Canada}

\vspace{2cm}


\begin{abstract}

We evaluate the coefficients of the leading poles of the complete two-loop 
quark self-energy $\Sigma(p)$ in the Coulomb gauge. Working in the framework of 
split dimensional regularization, with  complex regulating parameters 
$\sigma$ and $n/2-\sigma$ for the energy  and  space components of
the loop momentum, respectively, we find that 
 split dimensional regularization leads to well-defined two-loop 
integrals, and that the overall coefficient of the leading pole  term for $\Sigma(p)$
is strictly local. 
Extensive tables showing the pole parts of one- and two-loop Coulomb integrals
are given. 
We also comment on some general implications of split dimensional regularization, 
discussing in particular  the limit $\sigma\to 1/2$ and 
the subleading terms in the $\ep$-expansion of noncovariant integrals.

\end{abstract}

\end{center}
\normalsize

\end{titlepage}

\newpage

\section{Introduction}

Despite serious efforts during the past twenty years to place the Coulomb gauge
on the same rigorous footing as covariant gauges, we still have no consistent 
rules for renormalizing non-Abelian theories in this gauge. 
There can be little doubt, however, that the Coulomb gauge is 
superior to other gauges in at least two  respects, namely in  
the treatment of bound states and in the study of confinement in QCD. 
Since these topics play a  crucial role in our understanding of the
strong interactions, there is clearly a need to put  this gauge on a 
sound theoretical basis.
To gain a better understanding of the various advantages, as well as disadvantages, 
of the Coulomb gauge, we refer the reader to the vast literature on the subject. 
(For some recent references see, for example, refs.~\cite{lucha}--\cite{wilklev}.)\\
Unfortunately, progress in the Coulomb gauge
continues to be hampered by the operator ordering problem in the quantum 
Hamiltonian, as noted by Schwinger in 1962~\cite{schwinger}. 
The ordering problem was later re-examined by Christ and Lee who demonstrated 
that the quantum Hamiltonian differed from the classical Hamiltonian by special 
Coulomb interaction terms, labelled $(V_1+V_2)$~\cite{chlee}. A few years later,
Cheng and Tsai~\cite{chtsai1} pointed out that the $(V_1+V_2)$ -terms 
are equivalent to a distinct class of integrals, called {\it energy integrals}, 
which lead to two different types of divergences~\cite{chtsai1,taylor}: \\
(a) ordinary UV divergences associated with the structure of space-time, and \\
(b) divergences characteristic of the Coulomb gauge, arising from the integration 
over the energy variable $q_0$ in integrals of the form 
$$\int\limits_{-\infty}^{\infty} dq_0\,\frac{q_0^2}{q_0^2-\vec{q}^{\,2}+i\varepsilon}\;.$$ 
It is the divergences 
from such energy integrals that give rise to ambiguities in the Coulomb gauge. 
These ambiguities  come as no surprise, since the Coulomb-gauge condition 
\be
\vec\nabla\cdot \vec A(x)=0
\ee
 does not fix the gauge completely, but leaves a 
residual gauge freedom for gauge transformations $g(t)$ that do not depend on the
space coordinates $\vec{x}$. 
In 1987, Doust and Taylor~\cite{doust,doustay} came to the conclusion that 
standard dimensional regularization is incapable of regulating both types 
of divergences simultaneously\footnote{These authors assume that the energy 
component $q_0$ is one-dimensional, while $\vec{q}$ is $(n-1)$-dimensional.}. 

\medskip

Renormalization in the Coulomb gauge has recently 
been  examined by Baulieu and Zwanziger~\cite{bauzwa}, 
who treated the  Coulomb gauge as the  singular limit of the 
Landau-Coulomb interpolating gauge. Zwanziger also exploited the
Coulomb gauge in the  study  of confinement~\cite{zwanzig}. 
Employing a diagrammatic representation, he showed that 
the problematic energy integrals cancel, at least to  one-loop
order. Based on this observation, he assigned to these integrals  the 
value zero. However, it is not obvious that this  cancellation procedure
can also be carried out explicitly beyond one-loop order~\cite{taylor}. 

\medskip

Accordingly, it seems desirable to  regulate the Coulomb-gauge integrals on an
individual basis. 
To this effect, a novel technique called {\it split dimensional regularization}
was introduced by one of the authors~\cite{cou1}. The idea is to replace 
the measure $d^nq$ by
\be
d^{2(\sigma+\omega)}q=d^{2\sigma}q_0\,d^{2\omega}\vec{q} \;,\label{sdr}
\ee
where $\sigma$ and $\omega$ are understood as  parameters 
in the complex plane satisfying $\sigma+\omega=n/2$, 
and the limits $\sigma\to 1/2$ and $\omega\to 3/2$ are taken only at the 
end of  integration. One may think of split dimensional regularization
as a special form of dimensional regularization, the special feature 
being that the dimension of 
the energy component is explicitly specified to be non-integer.
Whereas in~\cite{zwanzig} the ambiguous energy integrals were {\it defined} 
to be zero, in the context of 
split dimensional regularization these  integrals turn out to be zero in a natural way. 

\medskip

To date, split dimensional regularization has been tested at one loop for the 
gluon self-energy~\cite{cou1}, and the quark self-energy and quark-gluon 
three-point function~\cite{cou2}. All Coulomb-gauge 
integrals appearing in these calculations are free of ambiguities and respect
 the appropriate Ward identities. \\
We note in passing that split dimensional regularization has also been applied 
in the context of non-relativistic QCD~\cite{griesshammer}.
 
\medskip

However, as already alluded to in ref.~\cite{cou2}, the real challenge 
comes from  energy integrals  at two loops and beyond~\cite{doustay}. 
In order to investigate the associated ambiguities, 
we have evaluated the contributions of the leading $1/\ep^2$ poles of the 
complete two-loop quark self-energy. Working in the framework of 
split dimensional regularization, we find that  all 
integrals  can be calculated consistently. \\
But, whereas the coefficients of the leading poles can be evaluated explicitly, 
the coefficients of the subleading poles are generally more difficult to 
compute, since many of the parameter integrals can no longer be expressed in 
closed form. For this reason, the subleading poles have not been evaluated 
in their entirety in this paper.

\medskip

It is also worth noting at this stage that the subleading poles 
exhibit the following interesting feature: 
 the regulators $\sigma$ and $\omega$, characterizing the time and space 
components, respectively, appear {\it independently} in the various 
$\Gamma$-functions,
a clear indication of the special role played by the time component in the 
Coulomb gauge. (See Section~\ref{noncosec} for more details.) 
By comparison, in the $\Gamma$-functions for the leading poles, the regulators 
$\sigma$ and $\omega$ always appear in the combination $\sigma+\omega=n/2$.

\medskip

The paper is organized as follows. In Section 2, we first discuss 
a specific integral for which standard dimensional regularization fails to regulate 
the divergence from the energy component, if the latter component is assumed to be 
one-dimensional, and then show how split 
dimensional regularization cures the ambiguity. 
In the second part of Section 2, we concentrate on  two-loop integrals, 
analyzing in particular the properties of the subleading poles. Then we 
comment on  some general properties of split dimensional regularization. 
We demonstrate that  the limit $\sigma\to 1/2$ 
{\it after} integration  is always well defined and discuss the implications of 
split dimensional regularization
for the subleading terms in the $\ep$-expansion of noncovariant integrals.
In Section 3, we outline the 
calculation of the leading divergence of the two-loop quark self-energy and
present our results. The highlights of this paper are summarized in Section 4. 
Appendix A contains results for the pole parts of one-loop integrals (for both 
integer and  non-integer powers of propagators), while Appendix B shows the
 results for the leading poles of several two-loop integrals. In Appendix C and D
we give general formulas for   Coulomb integrals at one and 
two loops in Feynman parameter space.

\section{Split dimensional regularization}

\subsection{Problems with standard dimensional regularization}\label{sec21}

The gluon propagator in the Coulomb gauge, given by~\cite{cou1}
\bea
G_{\mu\nu}^{ab}(q)&=&\frac{-i\delta^{ab}}{q^2+i\ep}\,d_{\mu\nu}^{\rm cou}(q)\;,\nonumber\\
d_{\mu\nu}^{\rm cou}(q)&=&g_{\mu\nu}+\frac{n^2}{\vec{q}^{\,2}}q_{\mu}q_{\nu}-
\frac{qn}{\vec{q}^{\,2}}\,(q_{\mu}n_{\nu}+n_{\mu}q_{\nu})\;,\label{dcou}\\
&&\nonumber\\
n_{\mu}&=&(1,0,0,0)\;,\nonumber
\eea
is seen to contain the  factor $1/\vec{q}^{\,2}$. 
While  most  Coulomb-gauge integrals can be computed consistently 
with  standard  dimensional regularization, it is no secret that the appearance
of the noncovariant factor $1/\vec{q}^{\,2}$ in the integrand may, under certain 
circumstances, lead to spurious singularities that require special attention. 
The presence of {\it two} or more of these factors\footnote{Note that in 
Abelian gauge theories two noncovariant factors containing 
the same loop momentum cannot occur if the number of external legs is $\le 3$, i.e.
in self-energy and vertex diagrams.},  
in combination with powers of $q_0$ in the numerator, 
 is particularly troublesome 
and can lead to the ambiguous energy integrals mentioned in Section 1.
To illustrate this point, consider the integral
\be
I_{\mu\nu}=\int d^n q\frac{q_{\mu}q_{\nu}}{q^2\,\vec{q}^{\,2}(\vec{p}-\vec{q})^2}
\;.
\ee
We shall first show that standard dimensional regularization fails to regulate the 
singularity in the energy component of $I_{\mu\nu}$ (i.e. in $I_{00}$), if 
 the energy component $q_0$ of a vector $q_{\mu}$ is assumed to be one-dimensional. 
After Feynman parametrization, we obtain
\be
I_{\mu\nu}=\Gamma(3)\int_0^1 dx\int_0^1dy\,y\int d^n q \frac{q_{\mu}q_{\nu}}{[qA^{-1}q-2q{\cal P}+M^2]^3}\;,
\ee
where
\be
A^{-1}=\left(\begin{array}{cc}xy & 0\\0& -{\bf 1}\end{array}\right)\; ;\;
{\cal P}=\left(\begin{array}{c} 0\\(1-y)\vec{p}\end{array}\right)\; ;\;
M^2=(1-y)\vec{p}^{\,2}\;.\label{Ama}
\ee
Integration over 
 $n=(4-2\ep)$-dimensional momentum space leads to
\bea
I_{\mu\nu}&=&i\pi^{\frac{n}{2}}\int_0^1 dx\int_0^1dy\,y |\mathrm{Det}A|^{\frac{1}{2}}\Bigl\{\Gamma(3-\frac{n}{2})\,({\cal P}A)_{\mu}({\cal P}A)_{\nu}\,[M^2-{\cal P}A{\cal P}]^{\frac{n}{2}-3}\nonumber\\
&&+\frac{1}{2}\Gamma(2-\frac{n}{2})\,A_{\mu\nu}\,[M^2-{\cal P}A{\cal P}]^{\frac{n}{2}-2}\Bigr\}\;.\label{detA}
\eea
The important term in Eq.~(\ref{detA}) is $\mathrm{Det}A$. If we assume that the 
energy component is one-dimensional, while the identity matrix contained in $A$
is defined in $(n-1)$ dimensions, we obtain 
$|\mathrm{Det}A|^{\frac{1}{2}}=(xy)^{-\frac{1}{2}}$, so that
\bea
I_{\mu\nu}&=&i\pi^{\frac{n}{2}}(\vec{p}^{\,2})^{-\ep}\int_0^1 dx\,x^{-\frac{1}{2}}\int_0^1dy\,y^{\frac{1}{2}}\nonumber\\
&&\Bigl\{\Gamma(1+\ep)(p_{\mu}-p_0n_{\mu})(p_{\nu}-p_0n_{\nu})\,y^{-1-\ep}(1-y)^{1-\ep}\nonumber\\
&&+\frac{1}{2}\Gamma(\ep)\,\left[\frac{1}{xy}\,n_{\mu}n_{\nu}+(g_{\mu\nu}-n_{\mu}n_{\nu})\right]\,y^{-\ep}(1-y)^{-\ep}\Bigr\}\;.\label{fail}
\eea
Thus, we see  that  the   ``energy component'' $I_{00}$
 is ill defined, since it leads to the parameter 
integral 
$$\int_0^1 dx\,x^{-\frac{3}{2}}\;.$$ 
The example above demonstrates that the application of standard dimensional 
regularization in the noncovariant Coulomb gauge leads to unavoidable 
difficulties, as long as the energy component $q_0$ is assumed to be one-dimensional.
This conclusion raises another question, namely, 
how is the dimensionality of $q_0$ defined in standard dimensional regularization?
The answer to this question is that in general, 
the dimension of $q_0$ need not to be specified. 
In covariant integrals, we could distribute arbitrary fractions of $\ep$ to the individual 
components of a Lorentz vector, as, for example, in
$$\int d^nq \to \int d^{\sigma_0}q_0\,d^{\sigma_1}q_1\,d^{\sigma_2}q_2\,d^{n-\sigma_0-\sigma_1-\sigma_2}q_3\;.$$
Due to covariance, the result will invariably depend only on $n$. However, in noncovariant 
gauges, the situation is quite different. 
In our example, 
we could as well have assumed that the identity matrix contained in $A$ is 
three-dimensional, thereby leading to an  
 energy component which is $(n-3)$-dimensional, so that
\be
|\mathrm{Det}A|^{\frac{1}{2}}=(xy)^{-\frac{n-3}{2}}=(xy)^{-\frac{1}{2}+\ep}\;.
\label{standard}
\ee
In that case, the $x$-integral in $I_{00}$ is given by 
$\int_0^1 dx\,x^{-\frac{3}{2}+\ep}$, which
 is well defined by invoking analytic continuation 
in the context of dimensional regularization. This is exactly the idea of split 
dimensional regularization. It asserts that in the noncovariant Coulomb gauge,
 the energy component 
$q_0$ of a Lorentz vector (as a loop momentum) cannot be treated as one-dimensional.
 Instead, a nonzero ``fraction'' of the regulator
$\ep$ in $n=4-2\ep$ has to be assigned to the energy component, for instance 
 by defining
 \bea
\sigma&=&\frac{1}{2}(1- \ep_{\sigma})\quad,\quad\ep_{\sigma}=c_{\sigma}\cdot \ep\label{cc}\;,\\
\omega&=&\frac{3}{2}-\ep\,(1-\frac{c_{\sigma}}{2})\;,\nonumber
\eea  
such that $\sigma+\omega=n/2$. The parameter $c_{\sigma}$ is arbitrary\footnote{Note that 
in \cite{heckathorn}, a special version of split dimensional regularization with 
$c_{\sigma}=1/2$ has been used.},
as long as
we do not set it to zero  before all parameter integrals have been carried
out. {\it After} integration, we can define the integral at $\sigma=1/2$ by analytic
continuation, invoking the same arguments as in standard dimensional 
regularization. \\ 
Thus, the technique of split dimensional regularization 
imposes the condition  that the energy 
component $q_0$ {\it must} 
be of  {\it non-integer} dimension. 
\\
Using split dimensional regularization, 
we find for the integral $I_{\mu\nu}$: 
\bea
I_{\mu\nu}&=&i\pi^{\frac{n}{2}}(\vec{p}^{\,2})^{-\ep}n_{\mu}n_{\nu}\,\Gamma(\ep)\cdot \sigma\int_0^1 dx\,x^{-1-\sigma}\int_0^1dy\,y^{-\sigma-\ep}(1-y)^{-\ep}\label{split}\\
&&+\; p_ip_j- \mbox{ and } \delta_{ij}- \mbox{ terms } \quad ; \quad i,j\in\{1,2,3\}\;,
\nonumber
\eea
where $(\sigma+\omega)$ has been replaced by $n/2=2-\ep$ whenever $\sigma, \omega$ 
occur in this particular combination. 
Contraction of  the result (\ref{split}) with $n^{\mu}n^{\nu}$  yields for the 
pole part of 
$I_{00}=\int d^n q\,q_0^2/\{q^2\,\vec{q}^{\,2}(\vec{p}-\vec{q})^2\}\;,$
\bea
\mathrm{div}\,I_{00}&=&\mathrm{div }\left\{-i\pi^{\frac{n}{2}}(\vec{p}^{\,2})^{-\ep}\Gamma(\ep)\,\mathrm{Beta}(1-\sigma-\ep,1-\ep)\right\}\label{beta}\\
&\stackrel{\sigma\to\frac{1}{2}}{=}& i\pi^2\frac{1}{\ep}\,(-2)\;.\nonumber
\eea
As a crosscheck, we may write
$$
\mathrm{div}\,I_{00}=\mathrm{div }\int d^n q\frac{q^2+\vec{q}^{\,2}}{q^2\,\vec{q}^{\,2}(\vec{p}-\vec{q})^2}=\mathrm{div }\int d^n q\frac{1}{q^2\,(\vec{p}-\vec{q})^2}
=i\pi^2\frac{1}{\ep}\,(-2)\;. 
$$
$$\mbox{Note that }\quad
\int d^{2\sigma} q_0\,\int\frac{d^{n-2\sigma}\vec{q}}{\vec{q}^{\,2}(\vec{p}-\vec{q})^2}=0
\hspace{7.2cm}$$
in the context of split dimensional regularization, by the same arguments as 
$\int d^n q=0$ in standard dimensional regularization.
The breaking of covariance is manifest in Eq.~(\ref{beta}) from the  function 
Beta$(1-\sigma-\ep,1-\ep)$,  which contains 
$\sigma$, rather than  $\sigma+\omega$, in the energy component of $I_{\mu\nu}$.
It is evident from this example that split dimensional regularization leads to 
well-defined parameter integrals.

\subsection{Subleading poles}\label{subleadsec}

In order to extract interesting  quantities such as the two-loop renormalization 
constant for the quark self-energy in the Coulomb gauge, it is of course 
necessary to determine not only the leading ($1/\ep^2$) poles, but also the 
subleading ($1/\ep$) poles of the two-loop integrals. Since
the expressions for some of the Coulomb  integrals 
occurring in the two-loop quark self-energy are quite involved 
(as already noted above),  the subleading poles of these integrals  
can no longer be extracted in the form of analytic functions. 
Instead, they have to be
expressed as combinations of infinite series and hypergeometric functions.
We shall illustrate the extraction of the subleading poles by means of two 
examples.
The first example consists of an integral $I_1$ where all integrations can be 
done in closed form: 
\bea
I_1&=&\frac{1}{i^2\pi^n}\int \frac{d^nk_1\,d^nk_2}{k_1^2(k_1-k_2)^2\vec{k}_2^{\,2}(\vec{p}-\vec{k}_1)^2}\nonumber\\
&=&-\frac{1}{i\pi^{\frac{n}{2}}}\int\frac{d^nk_1}{k_1^2(\vec{p}-\vec{k}_1)^2\,(\vec{k}_1^2)^{\ep}}\Gamma(\ep)\,\mathrm{Beta}(\omega-1,1-\ep)\nonumber\\
&=&(\vec{p}^{\,2})^{-2\ep}\,\Gamma(\ep)\Gamma(2\ep)\cdot G(\ep,\sigma,\omega)\;,\label{ex1}\\
&&\nonumber\\
G(\ep,\sigma,\omega)&=&\frac{\Gamma(1-\sigma)\Gamma^2(\omega-1)\Gamma(1-\ep)\Gamma(1-2\ep)}{\Gamma(1-\sigma+\ep)\Gamma(\omega-\ep)\Gamma(\omega-2\ep)}\;,\nonumber
\eea
where $\sigma+\omega=2-\ep$ has been used in Eq.~(\ref{ex1}). But note that $\sigma$ and $\omega$ also occur separately in $G(\ep,\sigma,\omega)$, not only in the 
combination $\sigma+\omega$. In order to extract the leading and subleading poles in $\ep$, 
 we use Eq.~(\ref{cc}) and expand\footnote{A factor $\Gamma^2(1+\ep)$ has been extracted in order to avoid the Euler $\gamma_E$ in the expanded result.}
in $\ep$. This yields 
\be
I_1=(\vec{p}^{\,2})^{-2\ep}\,\Gamma^2(1+\ep)\left\{\frac{2}{\ep^2}+
\frac{4}{\ep}\,\left[5-c_{\sigma}-2\log{(2)}\right] + \mbox{ finite}\right\}\;.
\label{com}
\ee
We will comment on the appearance of the parameter $c_{\sigma}$, in the subleading 
pole term, in Section~\ref{noncosec}. Note that the breaking of covariance is 
also manifest in the factor $(\vec{p}^{\,2})^{-2\ep}$.

\medskip

As a second example, consider the integral
\bea
I_2&=&\frac{1}{i^2\pi^n}\int\frac{d^nk_1\,d^nk_2}{k_1^2k_2^2(p-k_2)^2(k_1-k_2)^2\vec{k}_2^{\,2}}\nonumber\\
&=&(-1)^{-\ep}\Gamma(\ep)\,\mathrm{Beta}(1-\ep,1-\ep)\frac{1}{i\pi^{\frac{n}{2}}}
\int \frac{d^nk_2}{(k_2^2)^{1+\ep}(p-k_2)^2\vec{k}_2^{\,2}}\nonumber\\
&=&(-p^2)^{-1-2\ep}\Gamma(\ep)\,\mathrm{Beta}(1-\ep,1-\ep)\Gamma(-2\ep)\frac{\Gamma(1+2\ep)}{\Gamma(1-\ep)}\nonumber\\
&&\cdot \int_0^1 du\, u^{\omega-2}\,_2F_1(1+2\ep,-2\ep;1-\ep;B(u))\quad;\quad B(u)=\frac{p_0^2-u\vec{p}^{\,2}}{p^2}\;,\nonumber\\
&=&(-p^2)^{-1-2\ep}\Gamma^2(1+\ep)\frac{1}{\omega-1}\Bigl\{-\frac{1}{2\ep^2}\nonumber\\
&&-\frac{1}{\ep}\,\left[1-\sum_{m=1}^{\infty}\frac{1}{m}\left(\frac{p_0^2}{p^2}\right)^m\,_2F_1(-m,\omega-1;\omega;\frac{\vec{p}^{\,2}}{p_0^2})\right] + \mbox{ finite}\Bigr\}\;.
\eea
Using Eq.~(\ref{cc}), we obtain
\bea
I_2&=&(-p^2)^{-1-2\ep}\Gamma^2(1+\ep)\Big\{-\frac{1}{\ep^2}\nonumber\\
&&-\frac{1}{\ep}\,
\left[3-c_{\sigma}-\sum_{m=1}^{\infty}\frac{1}{m}\left(\frac{p_0^2}{p^2}\right)^m\,_2F_1(-m,\frac{1}{2};\frac{3}{2};\frac{\vec{p}^{\,2}}{p_0^2})\right] + \mbox{ finite}\Big\}\;.\label{ex2}
\eea
Again, we see that the breaking of Lorentz covariance manifests itself in terms 
such as  $\vec{p}^{\,2}/p_0^2$ and  in the fact that $c_{\sigma}$ appears in 
the subleading pole term.\\
It should be clear from the result  for the subleading poles 
of $I_2$ in Eq.~(\ref{ex2})
that a complete determination of the subleading poles in the two-loop 
quark self-energy is  beyond the scope of this paper. 
Afterall, $I_2$ is still  a relatively simple Coulomb  integral since it has 
only one noncovariant denominator. 

\subsection{Comment on noncovariance in the context of dimensional regularization}
\label{noncosec}

In the context of split dimensional regularization, 
as already explained in Section \ref{sec21}, the  dimension 
$d(q_0)$ of the energy component of an $n$-dimensional 
vector $q_{\mu}$ is given by  $d(q_0)=2\sigma$, 
where (cf. Eq.~(\ref{cc}))
\be
\sigma=\frac{1}{2}(1-\ep_{\sigma})\quad ,\quad\ep_{\sigma}\equiv c_{\sigma}\cdot\ep\;,
\label{csigma}
\ee
and $\ep_{\sigma}$ serves to regulate the spurious divergences in the 
energy component, inherent in the Coulomb gauge. 
The important point is that the limit $\ep_{\sigma}\to 0$ in the energy integrals
 {\it after} integration always exists. 
This means that the  ``singularities'' in the energy integrals, which would be
unregulated for $\ep_{\sigma}=0$, are really spurious. They do not show up 
as poles in $\ep_{\sigma}$. Only the ``usual'' UV and IR poles of 
Coulomb-gauge integrals appear as poles in $\ep=2-(\sigma+\omega)$, whereas 
the potential singularities  in the energy component  {\it never} show up 
as $\Gamma(\pm k\,(1/2-\sigma))$, but rather as $\Gamma(1-k-\sigma)$, 
where $k$ is a positive integer. Hence,  
 the limit $\ep_{\sigma}\to 0$ with $\ep$ fixed, i.e. the limit 
$c_{\sigma}\to 0$, indeed exists. This observation follows
from the expressions in Feynman parameter space for general Coulomb-gauge integrals 
given in Appendix \ref{APC} and \ref{APD}. First, consider a scalar one-loop 
integral with $r\:(r>0)$ covariant and $m$ noncovariant denominators 
in $n=2(\sigma+\omega)$ dimensions  (cf. Eq.~(\ref{a0})):
\bea
I_{r+m}(\alpha_j,\beta_l)&=&
\int\frac{d^nq}{i\pi^{\frac{n}{2}}}\prod_{j=1}^r\prod_{l=1}^m\,\frac{1}{(q-p_j)^{2\alpha_j}(\vec{q}-\vec{s}_l)^{2\beta_l}}\nonumber\\
&=&(-1)^{\sum_j\alpha_j}\frac{\Gamma(\lambda(\alpha,\beta)-\sigma-\omega)}{\prod_{j=1}^r\Gamma(\alpha_j)\prod_{l=1}^m\Gamma(\beta_l)}
\int_0^1\prod_{j=1}^rdx_j\,x_j^{\alpha_j-1}\left(\sum_{j=1}^r\,x_j\right)^{-\sigma}\nonumber\\
&&\prod_{l=1}^mdy_l\,y_l^{\beta_l-1}\,\delta(1-\sum_{j=1}^r\,x_j-\sum_{l=1}^m\,y_l)\nonumber\\
&&\left[-\sum_{j=1}^r\,x_j\,p_{0,j}^2+\frac{(\sum_{j=1}^r\,x_j\,p_{0,j})^2}{\sum_{j=1}^r\,x_j}+\sum_{j=1}^r\,x_j\vec{p_j}^2\right.\nonumber\\
&&\left.+\sum_{l=1}^m\,y_l\,\vec{s_l}^2-(\sum_{j=1}^r\,x_j\,\vec p_{j}+\sum_{l=1}^m\,y_l\,\vec{s_l})^2\right]^{\sigma+\omega-\lambda(\alpha,\beta)}\;,\label{a0generic}\\
\lambda(\alpha,\beta)&=&\sum\limits_{j=1}^r\alpha_j+\sum\limits_{l=1}^m\beta_l\;.\nonumber
\eea
Concerning the appearance of the parameters $\sigma$ and $\omega$, 
we see that the overall $\Gamma$-function  (which indicates the UV behaviour),
 as well as
 the term in square brackets containing the momentum dependence, always depend
  on the {\bf sum} $\sigma+\omega=n/2$.
Only the integration over the Feynman parameters
$x_j$ (which are associated with the covariant denominators) 
leads to the  isolated 
parameter $\sigma$, the reason being that the Feynman 
parameters $y_l$ never multiply a $q_0$ component. \\
Referring to  Eqs.~(\ref{tens})\,--\,(\ref{fa4}) in Appendix \ref{APC}, 
we observe that 
the most dangerous case for a potential singularity at $\sigma=1/2$ occurs when 
$r=1, l\ge 2$, and with factors of $q_0^b$ in the numerator, that is, 
in the following type of integral
(for simplicity, we assume $\beta_l =1$ for all $l$, 
since more general powers $\beta_l$ do not spoil the argument):
\bea
I_{1+m}^{(b)}(\alpha)&=&\int\frac{d^nq}{i\pi^{\frac{n}{2}}}\prod_{l=1}^m\,\frac{q_0^b}{(q-p)^{2\alpha}(\vec{q}-\vec{s}_l)^{2}}\nonumber\\
&=&(-1)^{\alpha}\frac{i^b}{\Gamma(\alpha)}
\int_0^1 dx\,x^{\alpha-1-\sigma}\int\prod_{l=1}^m dy_l\,\delta(1-x-\sum_{l=1}^m\,y_l)\nonumber\\
&&\int_0^{\infty}dz\,z^{\alpha+m-\frac{n}{2}-1}\exp\{-z\,[x\vec{p}^{\,2}+\sum_{l=1}^m\,y_l\,\vec{s_l}^2-(x\vec p+\sum_{l=1}^m\,y_l\,\vec{s_l})^2]\}\nonumber\\
&&\cdot\left(\frac{-\partial}{\partial a_4}\right)^b\exp\{-\Big(a_4p_4-\frac{1}{zx}\frac{a_4^2}{4}\Big)\}\,\Big|_{a_4=0}\;\;.\label{zbra}
\eea
Carrying out the derivative with respect to $a_4$, 
 we see that the resulting $x$-integral will be of the form 
\be
\int_0^1 dx\,x^{\alpha-1-\sigma-\gamma}\, f(x,y_l)\;,\;\mbox{ where }\quad\gamma=\left\{\begin{array}{ll}
b/2&\mbox{ for } b \mbox{ even}\\
(b-1)/2 &\mbox{ for } b \mbox{ odd}
\end{array}\right.
\;,\label{xscal}
\ee
such that $\gamma$ is always an integer. The function $f(x,y_l)$ corresponds to 
the term in square brackets in Eq.~(\ref{zbra}) and  will, after integration  
over $z$, always contain the sum
$\sigma+\omega$ in the exponent. Since $f(x,y_l)$  in general does not contribute  
  to the leading behaviour for $x\to 0$, 
it is irrelevant for our considerations\footnote{In the special cases where 
 $f(x,y_l)$ {\it does}  contribute to the leading behaviour for $x\to 0$ (for example, 
if $l=1, r=1, p=0$), it also 
provides a regulator $\ep$ for the exponent of $x$ and thus is harmless.}. \\
Note that values of $\alpha >1$ actually 
improve the behaviour for $x\to 0$ in Eq.~(\ref{xscal}). By contrast, it is
  the more severe {\it ultraviolet} 
behaviour, induced by powers of $q_0^b\;(b>0)$ in the numerator, 
that can  render the integral over $x$ singular unless the parameter $\sigma$ is 
   complex\footnote{By complex we mean that $\sigma$ 
has to be understood as a parameter in the complex plane in order to 
define the analytic continuation of the corresponding $\Gamma$-function.}.\\ 
Generalization of the above arguments  to $r>1$ is straightforward. If $r$ 
is the number of covariant propagators, $b$ the power of $q_0$ in the 
numerator and $l\ge 2$, then integration over one of the Feynman parameters 
$x_i$ requires a complex
value of $\sigma$, provided $r-b/2\le 0$ is fulfilled (or, respectively,
 $\sum_{j=1}^r\alpha_j-b/2\le 0$ for general propagator powers).
 \\
Summarizing, we conclude from the analysis above  that all parameter integrals 
are well defined if  $\sigma$ contains the regulator $\ep$, as in
$\sigma=(1-c_{\sigma}\,\ep)/2$, and that
the limit $c_{\sigma}\to 0$ {\it after} integration 
is non-singular,
since the  exponents $\alpha$ and $\gamma$ in Eq.~(\ref{xscal}) 
are never  half-integers.

\medskip

At two loops, the situation is similar, as may be seen from the expressions 
in Appendix \ref{APD}. If $c_1$ is the number of covariant denominators containing 
$q$, but not $k$, and $c_2$ is the number of covariant denominators containing 
$k$, but not $q$, and if, furthermore, $b_1$ and $b_2$ are the powers of $q_0$ and 
$k_0$ in the 
numerator, respectively, and $i_B$ is  defined as in Eq.~(\ref{ib}), 
then a non-integer dimension  of the energy components $q_0,k_0$  is mandatory 
in the following cases ($i\in\{1,2\}$):
\bea
c_i-b_i/2+i_B\le 0&&\mbox{ for }\; c_1 \mbox{ and } c_2\not =0\;,\nonumber\\
c_i-b_i/2\le 0&&\mbox{ for }\; i_B=1\mbox{ and }c_1\mbox{ or }c_2 =0\;.\label{2loocon}
\eea
As an example, consider the case $i_B=0, \lambda_1=1, \rho_1=1,  \lambda_i =0 $ 
for $i>1$, 
$\rho_j=0$ for $j>1$, and all $\alpha_l,\beta_u=1$ in Eqs.(\ref{Itwo})\,--\,(\ref{Iscal}). 
(More general propagator powers do not change the arguments.) In that case, we have
$$ {\cal M}_4=\left(\begin{array}{ll}
x_1&0\\
0&x_2\end{array}\right)\quad ;\quad (\mbox{Det}{\cal M}_4)^{-\frac{1}{2}}=(x_1\,x_2)^{-\sigma}\;.$$
Analogous to the one-loop case, additional powers of $q_0$ or $k_0$ in the numerator 
 lead to more negative powers of  $x_1$ or $x_2$, 
as can be seen from Eqs.~(\ref{nums}). For example,  two powers of   $q_0$ or $k_0$ in 
the numerator lead to terms such as 
\bea
(\mbox{Det}{\cal M}_4)^{-\frac{1}{2}}J_{l}[l_4^{(i)}l_4^{(j)}]
&\to&
 (\mbox{Det}{\cal M}_4)^{-\frac{1}{2}}\left({\cal M}_4^{-1}\right)^{(ij)}J_{l}[1]\nonumber\\
&\to& \delta_{ij}\int_0^1 dx_i\,x_i^{-1-\sigma}\quad ,\quad i,j\in\{1,2\}\;.\label{2losing}
\eea
Although higher powers of $q_0$ or $k_0$ in the numerator will lead to even 
more negative powers of 
$x_i$, such as $x_i^{-1-\sigma-\gamma}$, the parameter  $\gamma$  can never 
be half-integer. As a result,  
the parameter integral is well defined for complex values of $\sigma$,  
and the limit $\sigma\to1/2$ {\it after} integration exists due to the analytic 
properties of the  $\Gamma$-function. \\
These arguments and  the conclusion that $\gamma$  will never be  
half-integer, can be generalized to $L$ loops by using the same formalism 
as in Appendix \ref{APD}, with the objects $l_4,{\cal N}_4,\ldots$ and 
$\vec{l},\vec{\cal N},\ldots$ 
being  defined in $L\cdot 2\sigma$ and $L\cdot 2\omega$ dimensional space, 
respectively. 
Hence, it follows that a troublesome $\Gamma$-function of the form
$\Gamma(\pm k\,(1/2-\sigma))$ never arises.

\medskip

The arbitrariness in the choice of $c_{\sigma}$ is not problematic,
since it is just the noncovariant version of the arbitrariness in the definition 
of the subtraction scheme in the context of renormalization. 
To illustrate this point,
consider  a simple one-loop integral $T(p)$ in split dimensional
regularization:
\bea
T(p)&=&\int \frac{d^nq}{(2\pi)^n}\frac{1}{(p-q)^2\vec{q}^{\,2}}\nonumber\\
&=&\frac{1}{(2\pi)^n}\int d^{2\sigma}q_0\,d^{n-2\sigma}\vec{q}\,\frac{1}{(p-q)^2\vec{q}^{\,2}}\nonumber\\
&=&-\frac{i\pi^{\frac{n}{2}}}{(2\pi)^n}(\vec{p}^{\,2})^{\frac{n}{2}-2}\Gamma(2-\frac{n}{2})\,\mathrm{Beta}(\frac{n}{2}-1,\frac{n}{2}-1-\sigma)\;.\label{betasigma}
\eea
Using $n=4-2\,c_n\,\ep$, together with Eq.~(\ref{csigma}), and expanding
 in $\ep$, we arrive at
\be
T(p)=-2\,\frac{i}{(4\pi)^2}(\vec{p}^{\,2})^{-c_n\ep}\left\{\frac{1}{c_n\ep}+4-2\ln(2)-\frac{c_{\sigma}}{c_n}+\ln(4\pi)-\gamma_E\right\}\;.
\label{Tex}
\ee
The factor of $1/c_n$ in the pole part is irrelevant since it will be 
removed by renormalization. 
As the terms $\ln(4\pi)$ and $\gamma_E$, in the subleading part of the 
$\ep$-expansion, are 
always present as an artifact of standard dimensional regularization, 
the additional term $c_{\sigma}/c_n$ is just 
another  aspect of the  arbitrariness inherent in 
renormalization, caused by the breaking of covariance. 
The freedom in the 
definition of the finite part merely reflects the fact that  
renormalization schemes need to be specified.  
The procedure of dealing with  the parameter $c_\sigma$ during renormalization
(i.e., whether or not it is subtracted together with the terms 
$\ln(4\pi)-\gamma_E$) 
defines a specific ``noncovariant renormalization scheme'' in the context of 
split dimensional regularization.\\
Of course, the easiest option would be to set $c_\sigma$ to zero 
after having evaluated the integral, since, as explained above, 
the operation of 
 setting $c_\sigma=0$ {\it after} integration is well defined, 
and is the natural choice as it corresponds to 
 $d(q_0)=1$ and  $d(\vec{q})=n-1$  in standard dimensional regularization.
Nevertheless, it may be advantageous to keep the parameter $c_{\sigma}$ 
as a useful check when computing  gauge invariant quantities, 
since $c_{\sigma}$ obviously  has to cancel out in any physical quantity.

\section{Leading divergence of the two-loop quark self-energy}

\begin{figure}[h!]
\begin{center}
\begin{picture}(200,250)(0,-240)
\GlueArc(0,0)(16,0,180){2}{4}
\GlueArc(0,0)(28,0,180){3}{6}
\ArrowLine(-60,0)(-30,0)
\Line(-30,0)(60,0)
\Text(-57,-7)[]{$p$}
\Text(0,-23)[]{(a)}
\GlueArc(150,0)(25,0,180){3}{6}
\GlueArc(170,0)(25,0,180){3}{6}
\ArrowLine(100,0)(130,0)
\Line(130,0)(220,0)
\Text(103,-7)[]{$p$}
\Text(160,-23)[]{(b)}
\Gluon(0,-69)(0,-100){3}{3}
\GlueArc(0,-100)(28,0,180){3}{6}
\ArrowLine(-60,-100)(-30,-100)
\Line(-30,-100)(60,-100)
\Text(-57,-107)[]{$p$}
\Text(0,-123)[]{(c)}
\GlueArc(160,-70)(12,0,360){3}{8}
\Gluon(115,-100)(145,-70){3}{4}
\Gluon(172,-70)(205,-100){3}{4}
\ArrowLine(100,-100)(220,-100)
\Text(103,-107)[]{$p$}
\Text(160,-123)[]{(d)}
\ArrowArc(80,-170)(14,-40,320)
\Gluon(35,-200)(66,-170){3}{4}
\Gluon(94,-170)(125,-200){3}{4}
\ArrowLine(20,-200)(140,-200)
\Text(23,-207)[]{$p$}
\Text(80,-223)[]{(e)}
\end{picture}
\end{center}
\caption{Contributions to the two-loop quark self-energy}\label{fig1}
\end{figure}
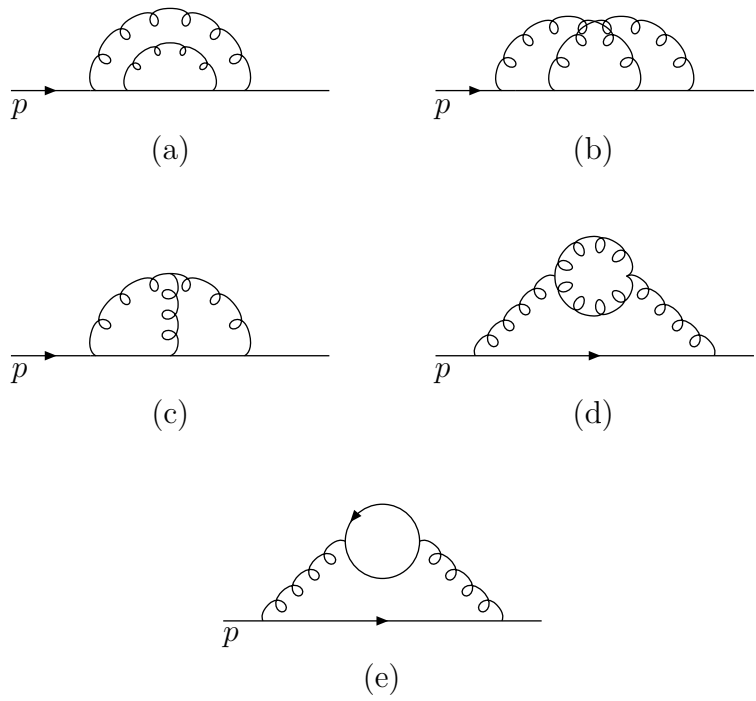

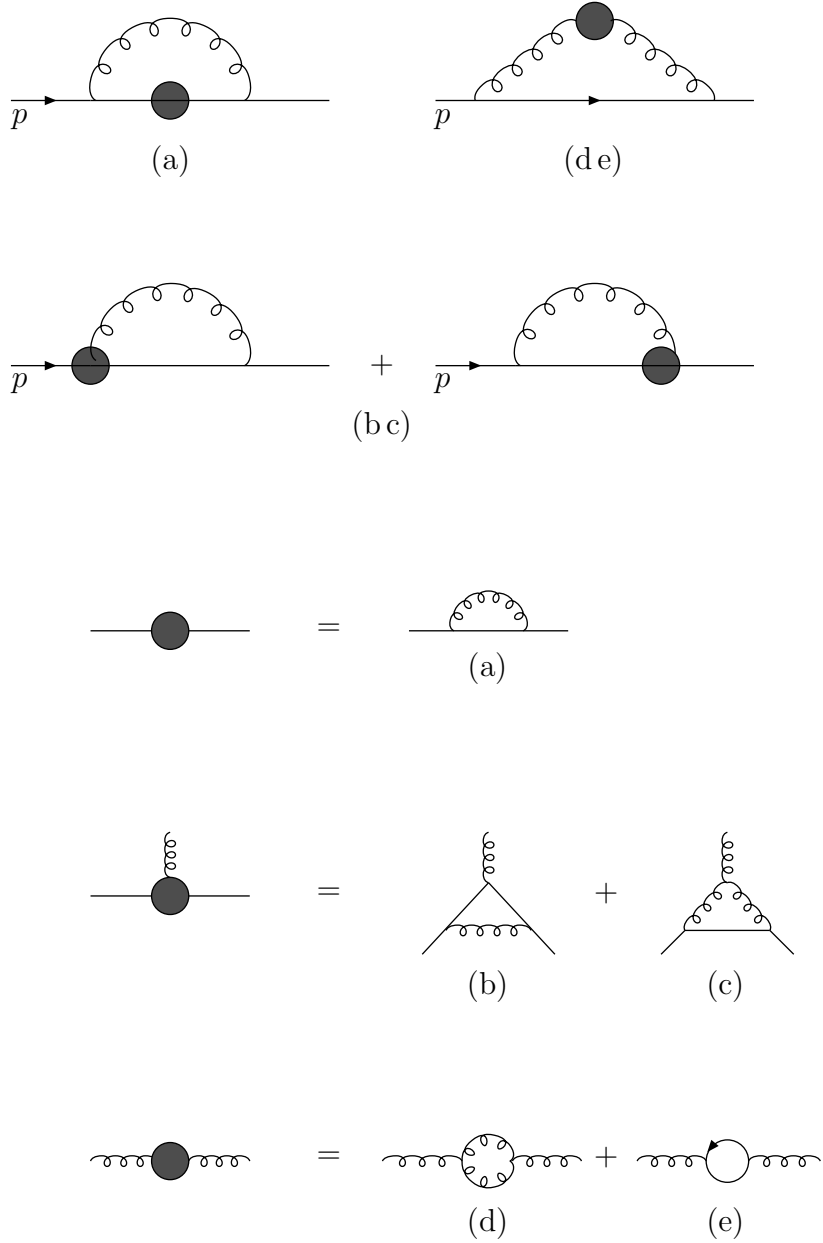
\begin{figure}[h] 
\begin{center} 
\begin{picture}(200,500)(0,-450)
\GCirc(0,0){7}{0.3}
\GlueArc(0,0)(28,0,180){3}{6}
\ArrowLine(-60,0)(-30,0)
\Line(-30,0)(60,0)
\Text(-57,-7)[]{$p$}
\Text(0,-23)[]{(a)}
\GCirc(160,30){7}{0.3}
\Gluon(115,0)(153,30){3}{4}
\Gluon(166,30)(205,0){3}{4}
\ArrowLine(100,0)(220,0)
\Text(103,-7)[]{$p$}
\Text(160,-23)[]{(d\,e)}
\GlueArc(160,-100)(28,0,180){3}{6}
\GCirc(185,-100){7}{0.3}
\ArrowLine(100,-100)(130,-100)
\Line(130,-100)(220,-100)
\Text(103,-107)[]{$p$}
\Text(80,-123)[]{(b\,c)}
\Text(80,-100)[]{+}
\GCirc(-30,-100){7}{0.3}
\GlueArc(0,-100)(28,0,176){3}{6}
\ArrowLine(-60,-100)(-30,-100)
\Line(-30,-100)(60,-100)
\Text(-57,-107)[]{$p$}
\Line(-30,-200)(-5,-200)
\GCirc(0,-200){7}{0.3}
\Line(30,-200)(7,-200)
\Text(60,-200)[]{=}
\Line(90,-200)(150,-200)
\GlueArc(120,-200)(13,0,180){2}{6}
\Text(120,-215)[]{(a)}
\Line(-30,-300)(-5,-300)
\GCirc(0,-300){7}{0.3}
\Line(30,-300)(7,-300)
\Gluon(0,-293)(0,-276){2}{3}
\Text(60,-300)[]{=}
\Line(95,-322)(120,-295)
\Line(120,-295)(145,-322)
\Gluon(120,-295)(120,-276){2}{3}
\Gluon(104,-313)(136,-313){2}{4}
\Text(120,-335)[]{(b)}
\Text(165,-300)[]{+}
\Line(185,-322)(194,-313)
\Line(226,-313)(235,-322)
\Gluon(194,-313)(210,-295){2}{3}
\Gluon(210,-295)(226,-313){2}{3}
\Gluon(210,-295)(210,-276){2}{3}
\Line(194,-313)(226,-313)
\Text(210,-335)[]{(c)}
\Gluon(-30,-400)(-5,-400){2}{3}
\GCirc(0,-400){7}{0.3}
\Gluon(7,-400)(30,-400){2}{3}
\Text(60,-400)[]{=}
\Gluon(80,-400)(110,-400){2}{3}
\GlueArc(120,-400)(8,0,360){2}{6}
\Gluon(129,-400)(155,-400){2}{3}
\Text(120,-425)[]{(d)}
\Text(165,-400)[]{+}
\Gluon(218,-400)(245,-400){2}{3}
\ArrowArc(210,-400)(8,-40,320)
\Gluon(176,-400)(202,-400){2}{3}
\Text(210,-425)[]{(e)}
\end{picture}
\end{center}
\caption{Counterterms for the two-loop quark self-energy. The grey circles 
denote the UV divergent part of the corresponding one-loop insertions given  
below.}  
\label{fig2}
\end{figure}

There are two methods of determining the overall divergence of the two-loop 
quark self-energy, namely by
\begin{enumerate}
\item direct calculation of the two-loop integrals corresponding to the 
graphs shown in Fig.~\ref{fig1};
\item computing the counterterm diagrams shown in Fig.~\ref{fig2} by
first extracting the UV-divergent parts of the corresponding one-loop subgraphs,  
and then calculating the divergence of the overall diagram containing these 
one-loop insertions. 
\end{enumerate}

We have used both of these methods in order to check the consistency of our 
results. 
Note that method (1) requires us to keep the full $\ep$-dependence from the 
first integration if the momentum integrations are performed sequentially 
(this procedure is sometimes called the ``nested method''~\cite{jimmy}).
Since it may not always be possible to maintain, without expansion, the full 
$\ep$-dependence, the majority of our  two-loop integrals has been 
computed by integrating over both loop momenta simultaneously 
(also referred to as the ``matrix method''~\cite{jimmy}), as 
outlined in Appendix \ref{APD}.  \\
It is  worth emphasizing that the (spurious) infrared divergences 
are more severe in the Coulomb gauge than they are, for example, in the 
Feynman gauge, the severity being caused by the noncovariant factor 
$1/\vec{q}^{\,2}$ in the gluon propagator. 
A partial list of the noncovariant integrals used in 
our calculation is given 
in the Tables in Appendix \ref{APA} and \ref{APB}.
The number of integrals per graph obtained by method (1) is typically of 
the order of a few hundred. 
The algebra has been performed by using the symbolic manipulation program 
FORM~\cite{form}.

\clearpage

Application of  method (2) requires insertion of  
the following one-loop  counterterms~\cite{cou1,cou2}
(cf. Fig.~\ref{fig2}): 
\bea
\Sigma(p)&=&i\,C_F\frac{\as}{4\pi}\,\frac{1}{\ep}\not p\;,\label{subgraphs}\\
\Gamma_{\mu}^a&=&T^a(\Gamma_{\mu}^{(b)}+\Gamma_{\mu}^{(c)})\;,\nonumber\\
\Gamma_{\mu}^{(b)}&=&i\,(C_F-\frac{N_c}{2})\frac{\as}{4\pi}\,\frac{1}{\ep}(-\gamma_{\mu})\;,\nonumber\\
\Gamma_{\mu}^{(c)}&=&i\,\frac{N_c}{2}\frac{\as}{4\pi}\,\frac{1}{\ep}(-\gamma_{\mu})\;,\nonumber\\
\Pi_{\mu\nu}^{G,ab}(q)&=&i\,N_c\delta^{ab}\frac{\as}{4\pi}\,\frac{1}{\ep}\,\Bigl[q^2g_{\mu\nu}-q_{\mu}q_{\nu}\nonumber\\
&&-\frac{4}{3}\frac{qn}{n^2}(q_{\mu}n_{\nu}+q_{\nu}n_{\mu})+\frac{8}{3}\frac{q^2}{n^2}n_{\mu}n_{\nu}\Bigr]\;,\nonumber\\
\Pi_{\mu\nu}^{F,ab}(q)&=&i\,T_f\delta^{ab}\frac{\as}{4\pi}\,\frac{1}{\ep}\,(-\frac{4}{3})(q^2\,g_{\mu\nu}-q_{\mu}q_{\nu})\;,\nonumber\\
&&\nonumber\\
C_F&=&(N_c^2-1)/(2N_c)\quad ;\quad T_f=T_R\cdot n_f=\frac{1}{2}n_f\;.\nonumber
\eea
Note that $\Sigma(p)$ and $\Gamma_{\mu}$ satisfy the Ward identity 
\be
(p^{\prime}-p)^{\mu}(\Gamma_{\mu}^{(b)}+\Gamma_{\mu}^{(c)})=\Sigma(p)-\Sigma(p^{\prime})\;,\label{ward}
\ee
and that the divergent part of the ghost contributions, 
discussed in ref.~\cite{cou1}, is zero. 

\clearpage

The two-loop corrections to the quark self-energy shown in Fig.~\ref{fig1} are 
given by the  expressions in Eqs.~(\ref{Ta}) to~(\ref{Te}), 
where we use the following short-hand notations: 
\bea
c_1&=&k_1^2+i\varepsilon \; ;\;c_3=(p-k_1)^2+i\varepsilon\;,\nonumber\\
c_2&=&k_2^2+i\varepsilon \; ;\;c_4=(p-k_2)^2+i\varepsilon\;,\nonumber\\
c_5&=&(k_1-k_2)^2+i\varepsilon\;,\label{shortcc}\\
\mbox{and}\hspace{3cm}&&\nonumber\\
V^{\mu_1\mu_2\mu_3}_{3g}(q_1,q_2,q_3)&=&
g^{\mu_1\mu_2}(q_1^{\mu_3}-q_2^{\mu_3})+g^{\mu_2\mu_3}(q_2^{\mu_1}-q_3^{\mu_1})+
g^{\mu_3\mu_1}(q_3^{\mu_2}-q_1^{\mu_2})\;,\nonumber\\
d_{\mu\nu}^{\rm cou}(q)&=&g_{\mu\nu}+\frac{n^2}{\vec{q}^{\,2}}q_{\mu}q_{\nu}-
\frac{qn}{\vec{q}^{\,2}}\,(q_{\mu}n_{\nu}+n_{\mu}q_{\nu})\;.
\eea
The ``rainbow graph'',  diagram (a), is given by
\begin{eqnarray}
\Sigma^{(a)}(p)&=&\frac{ig^4}{(2\pi)^{2n}}\,C_F^2\int d^nk_1\,d^nk_2\,
\frac{\gamma^{\alpha}\not k_1\gamma^{\mu}(\not k_1-\not k_2)\gamma^{\nu}\not k_1\gamma^{\beta}}{c_1^2c_2c_3c_5}\nonumber\\
&&\cdot\, d_{\alpha\beta}^{\rm cou}(p-k_1)\,d_{\mu\nu}^{\rm cou}(k_2)\;,\label{Ta}
\end{eqnarray}
while the graphs with the Abelian and non-Abelian vertex subgraphs have the form
\bea
\Sigma^{(b)}(p)&=&\frac{ig^4}{(2\pi)^{2n}}\,(C_F^2-\frac{C_FN_c}{2})\int d^nk_1\,d^nk_2\,
\frac{\gamma^{\alpha}\not k_1\gamma^{\mu}(\not k_1-\not k_2)\gamma^{\beta}(\not p-\not k_2)\gamma^{\nu}}{c_1c_2c_3c_4c_5}\nonumber\\
&&\cdot\, d_{\alpha\beta}^{\rm cou}(p-k_1)\,d_{\mu\nu}^{\rm cou}(k_2)\;,\nonumber\\
&&\nonumber\\
\Sigma^{(c)}(p)&=&\frac{ig^4}{(2\pi)^{2n}}\,\frac{C_FN_c}{2}\int d^nk_1\,d^nk_2\,
\frac{\gamma^{\mu}(\not p-\not k_1)\gamma^{\rho}(\not p-\not k_2)\gamma^{\beta}}{c_1c_2c_3c_4c_5}\nonumber\\
&&\cdot\, d_{\mu\nu}^{\rm cou}(k_1)\,d_{\lambda\rho}^{\rm cou}(k_1-k_2)\,d_{\alpha\beta}^{\rm cou}(k_2)\,V^{\nu\alpha\lambda}_{3g}(k_1,-k_2,k_2-k_1)\;.\nonumber
\eea
The gluon self-energy insertion reads 
\beas
\Pi_{ab}^{\mu\nu,G}(q)&=& \frac{g^2}{(2\pi)^n}\frac{N_c}{2}\delta_{ab}\int d^n k
\frac{d_{\alpha\beta}^{\rm cou}(k)\,d_{\lambda\rho}^{\rm cou}(k-q)}{k^2(k-q)^2}\nonumber\\
&&\cdot V^{\mu\alpha\lambda}_{3g}(q,-k,k-q)\,V^{\nu\beta\rho}_{3g}(-q,k,q-k)\;,\nonumber
\eeas
such that 
\be
\Sigma^{(d)}(p)=\frac{ig^2}{(2\pi)^n}\,T^aT^b\int d^n k_1\,\frac{\gamma^{\alpha}(\not p-\not k_1)\gamma^{\beta}}{c_1^2c_3}\,d_{\alpha\mu}^{\rm cou}(k_1)\,\Pi_{ab}^{\mu\nu,G}(k_1)\,d_{\nu\beta}^{\rm cou}(k_1)\;.\label{glu}
\ee
The fermion loop insertion in diagram (e) is trivial (it does not contain 
a gluon propagator in the inner loop), and is given by
\bea
\Pi_{ab}^{\mu\nu,F}(q)&=&-i\,T_f\delta_{ab}\frac{\as}{4\pi}(4\pi)^{\ep}\Gamma(\ep)\,\frac{\Gamma^2(2-\ep)}{\Gamma(4-2\ep)}\cdot 8\,(-q^2)^{-\ep}(q^2\,g^{\mu\nu}-q^{\mu}q^{\nu})\;,\nonumber\\
&&\nonumber\\
\Sigma^{(e)}(p)&=&\frac{ig^2}{(2\pi)^n}\,T^aT^b\int d^n k_1\,\frac{\gamma^{\alpha}(\not p-\not k_1)\gamma^{\beta}}{c_1^2c_3}\,d_{\alpha\mu}^{\rm cou}(k_1)\,\Pi_{ab}^{\mu\nu,F}(k_1)\,d_{\nu\beta}^{\rm cou}(k_1)\;.\nonumber\\
&&\label{Te}
\eea
We would like to point out  that $\Sigma^{(d)}$ and $\Sigma^{(e)}$ contain 
the double noncovariant 
denominator $1/(\vec{k}_1^{2})^2$ stemming from the two terms of $d_{\mu_1\mu_2}^{\rm cou}(k_1)$. In diagram (e), those double noncovariant denominators are reduced 
to 
single ones by corresponding terms in the numerator, whereas no such 
cancellation occurs in diagram (d). 
The  integrals  for graph (d) containing a factor $1/(\vec{k}_1^{2})^2$ are 
those given in Table~\ref{1loop3} for $\beta=2$. 

\medskip

After inserting the appropriate integrals and performing various  
crosschecks, as outlined above, 
we obtain the following results for the leading poles:
\bea
\Sigma^{(a)}(p)&=&i\,C_F^2\left(\frac{\as}{2\pi}\right)^2\,\frac{1}{\ep^2}
\left(-\frac{\not p}{4}\right)\;,\nonumber\\
\Sigma^{(b)}(p)&=&i\,(C_F^2-\frac{C_FN_c}{2})\left(\frac{\as}{2\pi}\right)^2\,\frac{1}{\ep^2}\,\frac{\not p}{4}\;,\nonumber\\
\Sigma^{(c)}(p)&=&i\,\frac{C_FN_c}{2}\left(\frac{\as}{2\pi}\right)^2\,\frac{1}{\ep^2}\,\frac{\not p}{4}\;,\nonumber\\
\Sigma^{(d)}(p)&=&i\,C_FN_c\left(\frac{\as}{2\pi}\right)^2\,\frac{1}{\ep^2}\,
\left\{-\frac{41}{36}\not p+\frac{8}{9}\frac{pn}{n^2}\not n\right\}\;,\nonumber\\
\Sigma^{(e)}(p)&=&i\,C_FT_f\left(\frac{\as}{2\pi}\right)^2\,\frac{1}{\ep^2}\,\left(-\frac{\not p}{3}\right)\;.\label{result}
\eea
Note that the Ward identity (\ref{ward}) manifests itself in the equation
\be
\Sigma^{(a)}+\Sigma^{(b)}+\Sigma^{(c)}=0\; .\label{w2lo}
\ee
We also recall that the one-loop gluon self-energy, $\Pi_{\mu\nu}^{G,ab}(q)$,  
is not transverse in the Coulomb gauge, 
since it satisfies the Ward identity~\cite{cou1}
\be
q^{\mu}\Pi_{\mu\nu}^{G,ab}(q)+(q^2\,g_{\mu\nu}-q_{\mu}q_{\nu})H^{ab\mu}=0\; ,
\ee
where  $H^{ab\mu}$ is a non-vanishing ghost term.
It should  come as no surprise, therefore, that the final expression for 
$\Sigma^{(d)}$ in Eqs.~(\ref{result}) still 
contains the vector $n_{\mu}$. \\
Finally, we should like to draw the reader's attention to the complete absence of 
nonlocal factors in any of the leading-pole expressions, Eqs.~(\ref{result}).
The absence of 
terms such as  $p_0^2/\vec{p}^{\,2}, \vec{p}^{\,2}/p^2 $, etc.,
which implies the cancellation of all spurious infrared divergences, 
is certainly remarkable in view of the ubiquitous appearance of nonlocal 
terms  at intermediate stages of the calculation. 
Of course, the trend was already set at the one-loop level, where the divergent 
parts of both the gluon self-energy and the quark self-energy were shown
to be {\it local}~\cite{cou1,cou2}.

\section{Conclusion}

In this paper we have  tested the technique of split dimensional regularization 
 by calculating the leading divergence of the two-loop quark
self-energy $\Sigma(p)$ in the Coulomb gauge. We find that the application of
split dimensional regularization enables us to compute 
 all two-loop integrals  consistently, and that
 the final expressions for the leading poles of each graph are {\it local}, 
despite the presence of  nonlocal terms  at intermediate steps of the 
calculation. 
\\
The leading pole parts of  noncovariant  integrals at one  and two loops, 
used in the present calculation, can be found in Tables \ref{1loop1}\,--\,\ref{2loop2}
(Appendix \ref{APA}, \ref{APB}). 
We have also derived general expressions, in Feynman parameter space, for both 
one- and two-loop Coulomb-gauge integrals  
(Appendix \ref{APC}, \ref{APD}).\\
The latter expressions are useful in analyzing 
 some general properties of the technique of split 
dimensional regularization. First of all, these expressions  allow us 
to identify the class
of integrals for which a complex regulator of the form 
$\sigma=(1-c_{\sigma} \ep)/2$ is mandatory in order that these integrals be at 
all well defined. Furthermore, it turned out  that  the 
{\it leading} pole of a typical Coulomb-gauge integral, evaluated with 
split dimensional regularization,  depends only  on the {\bf sum} of the regulating 
parameters, i.e. on 
$\sigma+\omega=n/2$,  whereas the {\it subleading} pole  generally contains the
 parameter $c_{\sigma}$. The latter  stems from the isolated appearance of 
the regulator $\sigma$ and  is  a direct consequence 
of breaking  covariance.
Finally, we have demonstrated that  the limit $c_{\sigma} \to 0$ {\it after}
integration  always  leads to non-singular expressions,  thus establishing  
the  consistency of split dimensional regularization in general.

\newpage

{\bf\large Acknowledgements}

\medskip

We should like to thank Jimmy Williams for several discussions.  
This research was supported in parts by the Natural Sciences and Engineering 
Research Council of Canada under Grant No. A8063.

\appendix
\section{Divergent parts of one-loop integrals}\label{APA}

In the following tables we give the results in Minkowski space 
for the  poles of 
the one-loop integrals entering in the calculation. All tables have to be 
read as follows: The result for the pole part of an integral 
$$\frac{1}{i\pi^2}\int d^{2(\sigma+\omega)}q\;\frac{A}{B}$$ 
is listed such that the denominator $B$ is given in the first row and the 
corresponding  numerators $A$ in the first column. All entries are 
implicitly multiplied by $\Gamma(\ep)$. Note  that the poles 
which are of an infrared nature have a 
negative mass dimension (for example, a coefficient 
proportional to $1/(p^2)^a$, or $1/(\vec{p}^{\,2})^a$, where $a>0)$.
The symbol ``--'' means that the corresponding integral does not occur 
in our calculation and hence has not been computed.

\begin{table}[h!]
\begin{center}
\begin{tabular}{|l||c|c|c|}
\hline
&&&\\
$\qquad B$&$q^2(p-q)^2\vec{q}^{\,2}(\vec{p}-\vec{q})^2$&$q^2(p-q)^2(k-q)^2\vec{q}^{\,2}$&$(p-q)^2(k-q)^2\vec{q}^{\,2}$\\
$A$&&&\\
\hline
&&&\\
1&$4/(p^2\vec{p}^{\,2})$&$2/(p^2k^2)$&0\\
$q_0$&$2p_0/(p^2\vec{p}^{\,2})$&0&0\\
$q_i$&$2p_i/(p^2\vec{p}^{\,2})$&0&0\\
$q_iq_0$&$2p_ip_0/(p^2\vec{p}^{\,2})$&0&0\\
$q_0^2$&$2p_0^2/(p^2\vec{p}^{\,2})$&0&$-1$\\
$q_iq_j$&$2p_ip_j/(p^2\vec{p}^{\,2})$&0&$1/3\,\delta_{ij}$\\
&&&\\
\hline
\end{tabular}
\end{center}
\caption{Pole terms of one-loop integrals in Minkowski space.}
\label{1loop1} 
\end{table}

\begin{table}[h!]
\lspage{
\begin{center}
\begin{tabular}{|l||c|c|c||c|c|c|}
\hline
&\multicolumn{3}{c||}{}&\multicolumn{3}{c|}{}\\
$\qquad B$&\multicolumn{3}{c||}{$q^{2\alpha}(\vec{p}-\vec{q})^2$}&\multicolumn{3}{c|}{$q^{2\alpha}(p-q)^2\vec{q}^{\,2}$}\\
$A$&\multicolumn{3}{c||}{}&\multicolumn{3}{c|}{}\\
\hline
$\alpha$&1&$1+\ep$&$2+\ep$&1&$1+\ep$&$2+\ep$\\
\hline
&&&&&&\\
1&$-2$&$-1$&$-1/(2p^2)$&$2/p^2$&$1/p^2$&$[1+4\vec{p}^{\,2}/(3p^2)]/p^4$\\
$q_0$&0&0&0&0&0&$p_0/p^4$\\
$q_i$&$-4\,p_i/3$&$-2\,p_i/3$&0&0&0&$p_i/(3p^4)$\\
$q_iq_0$&0&0&0&0&0&0\\
$q_0^2$&$-2\,\vec{p}^{\,2}/3$&$-\vec{p}^{\,2}/3$&$-1/2$&$-1$&$-1/2$&$1/(2p^2)$\\
$\vec{q}^{\,2}$&$-2\,\vec{p}^{\,2}/3$&$-\vec{p}^{\,2}/3$&$1/2$&$1$&$1/2$&$-1/(2p^2)$\\
$q_iq_j$&$-16/15\,p_ip_j+$&$-8/15\,p_ip_j+$&$1/6\,\delta_{ij}$&$1/3\,\delta_{ij}$&$1/6\,\delta_{ij}$&--\\
&$2/15\,\vec{p}^{\,2}\,\delta_{ij}$&$1/15\vec{p}^{\,2}\,\delta_{ij}$&&&&\\
&&&&&&\\
\hline
\hline
&\multicolumn{3}{c||}{}&\multicolumn{3}{c|}{}\\
&\multicolumn{3}{c||}{$q^{2\alpha}\vec{q}^{\,2}(\vec{p}-\vec{q})^2$}&\multicolumn{3}{c|}{$q^{2\alpha}(p-q)^2$}\\
&\multicolumn{3}{c||}{}&\multicolumn{3}{c|}{}\\
\hline
$\alpha$&1&$1+\ep$&$2+\ep$&1&$1+\ep$&$2+\ep$\\
\hline
&&&&&&\\
1&$2/\vec{p}^{\,2}$&$1/\vec{p}^{\,2}$&&1&1/2&$-1/(2p^2)$\\
$q_0$&0&0&&$p_0/2$&$p_0/4$&0\\
$q_i$&0&0&&$p_i/2$&$p_i/4$&0\\
$q_iq_0$&0&0&&$p_ip_0/3$&$p_ip_0/6$&0\\
$q_0^2$&$-2$&$-1$&&$p_0^2/3-p^2/12$&$p_0^2/6-p^2/24$&1/8\\
$\vec{q}^{\,2}$&$-2$&$-1$&&$p^2/4+\vec{p}^{\,2}/3$&$p^2/8+\vec{p}^{\,2}/6$&$-3/8$\\
$q_iq_j$&$-2/3\,\delta_{ij}$&$-1/3\,\delta_{ij}$&&$p^2/12\,\delta_{ij}+p_ip_j/3$&$p^2/24\,\delta_{ij}+p_ip_j/6$&$-1/8\,\delta_{ij}$\\
&&&&&&\\
\hline
\end{tabular}
\end{center}
\caption{Pole terms of one-loop integrals. 
All entries have to be multiplied by the overall factor $\Gamma(\ep)$.}
\label{1loop2} 
}
\end{table}

\begin{table}[h!]
\begin{center}
\begin{tabular}{|l||c|c|c||c|c||c|c|}
\hline
&\multicolumn{3}{c||}{}&\multicolumn{2}{c||}{}&\multicolumn{2}{c|}{}\\
$\quad \;B$&\multicolumn{3}{c||}{$(p-q)^{2\alpha}\vec{q}^{\,2\beta}$}&\multicolumn{2}{c||}{$q^{2\alpha}(p-q)^{2}\vec{q}^{\,2\beta}$}&\multicolumn{2}{c|}{$q^{2}(\vec{p}-\vec{q})^{2\alpha}\vec{q}^{\,2\beta}$}\\
$A$&\multicolumn{3}{c||}{}&\multicolumn{2}{c||}{}&\multicolumn{2}{c|}{}\\
\hline
$\alpha$&1&1&1&$\ep$&$1+\ep$&$\ep$&1\\
$\beta$&$\ep$&$1+\ep$&2&2&2&1&$\ep$\\
\hline
&&&&&&&\\
1&0&$-1$&0&0&$[4+16\,\vec{p}^{\,2}/(3p^2)]/p^4$&1&$-1$\\
$q_0$&0&$-p_0$&0&0&$2p_0/p^4$&0&0\\
$q_i$&0&$-p_i/3$&0&0&--&0&$-2p_i/3$\\
$q_0^2$&--&$p^2/3-4p_0^2/3$&$-2$&--&--&0&$-2\vec{p}^{\,2}/15$\\
$q_0^3$&--&--&$-6p_0$&--&--&0&0\\
$q_0^4$&--&--&$2p^2-14p_0^2$&--&--&--&--\\
&&&&&&&\\
\hline
\end{tabular}
\end{center}
\caption{Pole terms of one-loop integrals with non-integer denominator powers.}\label{1loop3} 
\end{table}

\clearpage

\section{Leading poles of two-loop integrals}\label{APB}

The integrals listed are of the form 
$$\frac{1}{i^2\pi^4}\int d^{2(\sigma+\omega)}k_1\,d^{2(\sigma+\omega)}k_2\,\;\frac{A}{B}\;,$$
where  the denominator $B$ is given in the first row and the corresponding numerators $A$ in the first column. For the denominators, the short-hand notations defined in Eq.~(\ref{shortcc}), as well as $k_3=p-k_1$, have been used. 
Note that $k_0=k^0$, $p_0=p^0$.
All entries are implicitly multiplied by $1/\ep^2$. 

\begin{table}[h!]
\begin{center}
\begin{tabular}{|l||c|c|c|c|c|}
\hline
&&&&&\\
$\qquad\qquad B$&$c_1c_2c_3c_4\vec{k}_2^{\,2}\vec{k}_3^{\,2}$&$c_2c_3c_5\vec{k}_2^{\,2}\vec{k}_3^{\,2}$&$c_1c_3c_5\vec{k}_2^{\,2}\vec{k}_3^{\,2}$&$c_1c_2c_5\vec{k}_2^{\,2}\vec{k}_3^{\,2}$&$c_1c_2c_4\vec{k}_2^{\,2}\vec{k}_3^{\,2}$\\
$A$&&&&&\\
\hline
&&&&&\\
1&$4/p^2$&$4/p^2$&$-4/p^2$&$-1/\vec{p}^{\,2}$&$-4/p^2$\\
$k_2^0k_3^0$&0&0&$-1$&0&0\\
$(pk_2)k_2^0$&$-2p^0/p^2$&$-p^0$&--&$p^0$&$2p^0$\\
$(pk_2)k_2^0k_3^0$&0&0&--&$p_0^2$&$2p_0^2$\\
$(k_2k_3)k_2^0k_3^0$&1&0&--&--&--\\
&&&&&\\
\hline
\hline
&&&&&\\
&$c_3c_5\vec{k}_2^{\,2}\vec{k}_3^{\,2}$&$c_1c_4\vec{k}_2^{\,2}\vec{k}_3^{\,2}$&$c_1c_5\vec{k}_2^{\,2}\vec{k}_3^{\,2}$&&\\
&&&&&\\
\hline
&&&&&\\
1&$-2$&4&2&&\\
$k_2^0$&$-2p^0$&$4p^0$&0&&\\
$k_3^0$&0&$4p^0$&$2p^0$&&\\
$k_2^0k_3^0$&0&$4p_0^2$&$-2/3\,\vec{p}^{\,2}$&&\\
$pk_3$&0&$4/3(p^2+2p_0^2)$&$2/3(p^2+2p_0^2)$&&\\
&&&&&\\
\hline
\end{tabular}
\end{center}
\caption{Leading pole terms of two-loop integrals with two noncovariant denominators. }\label{2loop1} 
\end{table}

\begin{table}[h!]
\begin{center}
\begin{tabular}{|l||c|c|c|c|}
\hline
&&&&\\
$\qquad\quad B$&$c_1c_2c_3c_4\vec{k}_2^{\,2}$&$c_1c_2c_3c_5\vec{k}_2^{\,2}$&$c_1c_2c_4c_5\vec{k}_2^{\,2}$&$c_2c_3c_4c_5\vec{k}_2^{\,2}$\\
$A$&&&&\\
\hline
&&&&\\
1&$2/p^2$&$-1/p^2$&$1/p^2$&$2/p^2$\\
$k_2^0$&0&0&0&0\\
$k_2^0k_3^0$&0&0&1/4&1/4\\
$(pk_2)k_2^0$&$-p^0$&$-p^0/2$&$-p^0/2$&$-p^0/2$\\
$(pk_3)k_2^0$&0&0&$p^0/4$&$p^0/4$\\
$(k_2k_3)k_2^0$&$-p^0/2$&--&--&--\\
&&&&\\
\hline
\hline
&&&&\\
&$c_2c_3c_5\vec{k}_2^{\,2}$&$c_1c_3c_4\vec{k}_2^{\,2}$&$c_3c_4c_5\vec{k}_2^{\,2}$&\\
&&&&\\
\hline
&&&&\\
1&1&$-2$&$-1$&\\
$k_2^0$&0&$-2p^0$&$-p^0$&\\
$k_2^0k_3^0$&0&$-p_0^2$&$1/6\,\vec{p}^{\,2}$&\\
$pk_2$&0&$-2p^2/3-4p_0^2/3$&$-p^2/3-2p_0^2/3$&\\
$pk_3$&$p^2/2$&$-p^2$&$\vec{p}^{\,2}/3$&\\
&&&&\\
\hline
\end{tabular}
\end{center}
\caption{Leading pole terms of two-loop integrals with one noncovariant denominator. An overall factor $1/\ep^2$ is implicit. }\label{2loop2} 
\end{table}

\section{General formula for one-loop integrals in the Coulomb gauge}\label{APC}

Here, we give the expression in Feynman parameter space for scalar
one-loop integrals in the Coulomb gauge with an arbitrary number and arbitrary 
powers of propagators.
We also outline how to obtain expressions in Feynman parameter space for 
the tensor integrals. 
\\
Consider a scalar integral with $r$ covariant
 and $m$ noncovariant denominators in $n$ dimensions:
\bea
I_{r+m}(\alpha_j,\beta_l)&=&\int\frac{d^nq}{i\pi^{\frac{n}{2}}}\prod_{j=1}^r\prod_{l=1}^m\,\frac{1}{B_j^{\alpha_j}C_l^{\beta_l}}\;,\label{Inm}\\
B_j&=&(q-p_j)^2\quad ,\quad C_l=(\vec q-\vec s_l)^2\;.\nonumber
\eea
Going to Euclidean space and  introducing Feynman parameters in the usual way, 
we find that 
the integral $I_{r+m}(\alpha_j,\beta_l)$ is given by
\beas
I_{r+m}(\alpha_j,\beta_l)&=&(-1)^{\sum_j\alpha_j}\int\frac{d^nq_E}{\pi^{\frac{n}{2}}}\prod_{j=1}^r\prod_{l=1}^m\,\frac{1}{B_{E,j}^{\alpha_j}C_l^{\beta_l}}\\
&=&(-1)^{\sum_j\alpha_j}\frac{\Gamma(\lambda(\alpha,\beta))}{\prod_{j=1}^r\Gamma(\alpha_j)\prod_{l=1}^m\Gamma(\beta_l)}
\int_0^1\prod_{j=1}^rdx_j\,x_j^{\alpha_j-1}\\
&&\prod_{l=1}^mdy_l\,y_l^{\beta_l-1}\,\delta(1-\sum_{j=1}^r\,x_j-\sum_{l=1}^m\,y_l)
\int\frac{d^nq_E}{\pi^{\frac{n}{2}}}\;{\cal D}^{-\lambda(\alpha,\beta) }\;,\\
&&\\
\mbox{where}&&\\
q_E^2&=&q_4^2+\vec{q}^{\,2}\quad ,\quad \lambda(\alpha,\beta)=\sum\limits_{j=1}^r\alpha_j+\sum\limits_{l=1}^m\beta_l\;,\\
&&\\
{\cal D}&=&\vec{q}^{\,2}+\sum_{j=1}^r\,x_j\,q_4^2-2q_4\sum_{j=1}^r\,x_j\,p_{4,j}-2\vec q\,(\sum_{j=1}^r\,x_j\,\vec p_{j}+\sum_{l=1}^m\,y_l\,\vec s_l)\\
&&+\sum_{j=1}^r\,x_j\,p_j^2+\sum_{l=1}^m\,y_l\,\vec{s_l}^2\;.
\eeas
 In order to obtain a formula for general tensor integrals, we first define
\be
I_{r+m}(a,\alpha_j,\beta_l)=\int\frac{d^nq}{i\pi^{\frac{n}{2}}}\prod_{j=1}^r\prod_{l=1}^m\,\frac{{\bf e}^{-a\cdot q}}{B_j^{\alpha_j}C_l^{\beta_l}}\;,
\ee
where $a_{\mu}$ is an arbitrary Lorentz vector in Euclidean space, and then
 derive the tensor integrals by differentiation with respect to $a$:
\begin{eqnarray}
I_{r+m}^{\mu_1\ldots\mu_s}(\alpha_j,\beta_l)&=&\int\frac{d^n q}{i\pi^{\frac{n}{2}}}\prod_{j=1}^r\prod_{l=1}^m\,\frac{q^{\mu_1}\ldots q^{\mu_s}}{B_j^{\alpha_j}C_l^{\beta_l}}\nonumber\\
&=&
(-1)^s\frac{\partial}{\partial a_{\mu_1}}\cdots \frac{\partial}{\partial a_{\mu_s}}I_{r+m}(a,\alpha_j,\beta_l)\Big|_{a=0}\;.\label{tens}
\end{eqnarray}
Carrying out the momentum integration in $I_{r+m}(a,\alpha_j,\beta_l)$, we obtain
\bea
I_{r+m}(a,\alpha_j,\beta_l)&=&\frac{(-1)^{\sum_j\alpha_j}}{\prod_{j=1}^r\Gamma(\alpha_j)\prod_{l=1}^m\Gamma(\beta_l)}
\int_0^1\prod_{j=1}^rdx_j\,x_j^{\alpha_j-1}\left(\sum_{j=1}^r\,x_j\right)^{-\sigma}\nonumber\\
&&\prod_{l=1}^mdy_l\,y_l^{\beta_l-1}\,\delta(1-\sum_{j=1}^r\,x_j-\sum_{l=1}^m\,y_l)\int_0^{\infty}dz\,z^{\lambda(\alpha,\beta)-\sigma-\omega-1}\nonumber\\
&&\exp\Big\{-z\big[\sum_{j=1}^r\,x_j\,p_{4,j}^2-\frac{(\sum_{j=1}^r\,x_j\,p_{4,j})^2}{\sum_{j=1}^r\,x_j}+\sum_{j=1}^r\,x_j\vec{p_j}^2\nonumber\\
&&+\sum_{l=1}^m\,y_l\,\vec{s_l}^2-(\sum_{j=1}^r\,x_j\,\vec p_{j}+\sum_{l=1}^m\,y_l\,\vec s_l)^2\big]\Big\}\nonumber\\
&&\cdot \exp\{-f(a_4)\}\cdot \exp\{-g(\vec a)\}\;,\label{resinm}\\
&&\nonumber\\
f(a_4)&=&\frac{1}{\sum_{j=1}^r\,x_j}\left(a_4\sum_{j=1}^r\,x_j\,p_{4,j}-\frac{1}{z}\frac{a_4^2}{4}\right)\;,\label{fa4}\\
g(\vec a)&=&\vec a\left(\sum_{j=1}^r\,x_j\,\vec p_{j}+\sum_{l=1}^m\,y_l\,\vec s_l\right)-\frac{1}{z}\frac{\vec a^2}{4}\;.\label{gav}
\eea
Note that $f(a_4)$ and $g(\vec a)$ depend on $z$, leading to more severe 
UV divergences for higher rank tensor integrals. \\
For a scalar integral, we can immediately set $a=0$, carry out 
the $z-$integration and go back to Minkowski space to get
\bea
I_{r+m}(\alpha_j,\beta_l)
&=&(-1)^{\sum_j\alpha_j}\frac{\Gamma(\lambda(\alpha,\beta)-\sigma-\omega)}{\prod_{j=1}^r\Gamma(\alpha_j)\prod_{l=1}^m\Gamma(\beta_l)}
\int_0^1\prod_{j=1}^rdx_j\,x_j^{\alpha_j-1}\left(\sum_{j=1}^r\,x_j\right)^{-\sigma}\nonumber\\
&&\prod_{l=1}^mdy_l\,y_l^{\beta_l-1}\,\delta(1-\sum_{j=1}^r\,x_j-\sum_{l=1}^m\,y_l)\nonumber\\
&&\Big[-\sum_{j=1}^r\,x_j\,p_{0,j}^2+\frac{(\sum_{j=1}^r\,x_j\,p_{0,j})^2}{\sum_{j=1}^r\,x_j}+\sum_{j=1}^r\,x_j\vec{p_j}^2\nonumber\\
&&+\sum_{l=1}^m\,y_l\,\vec{s_l}^2-(\sum_{j=1}^r\,x_j\,\vec p_{j}+\sum_{l=1}^m\,y_l\,\vec{s_l})^2\Big]^{\sigma+\omega-\lambda(\alpha,\beta)}\;.\label{a0}
\eea

\section{General formula for two-loop Coulomb gauge integrals}\label{APD}

A general two-loop integral in the Coulomb gauge, with $r+m+1$ covariant and $a+b+1$ noncovariant denominators, is of the form
\bea
J_{r,m,a,b}&=&\int\frac{d^n q\,d^nk}{i^2\pi^{n}}\frac{\prod_{i=1}^r\prod_{j=r+1}^{r+m}\prod_{l=1}^a\prod_{u=a+1}^{a+b}}{(q-k)^2(\vec{q}-\vec{k})^2(q-p_i)^{2\lambda_i}(k-p_j)^{2\rho_j}(\vec{q}-\vec{s}_l)^{2\alpha_l}
(\vec{k}-\vec{s}_u)^{2\beta_u}}\;.\nonumber\\
&&\label{Itwo}
\eea
The general exponents $\lambda_i,\rho_j,\alpha_l$ and $\beta_u$ have been 
introduced in order to account for certain denominators such as 
$\left((q-p)^2\right)^2$, occurring for 
example in the gluon self-energy correction. \\
Going to Euclidean space and introducing Feynman parameters $x_j$ 
for the covariant denominators and $y_l$ for the noncovariant denominators, 
where $x_0$ and $y_0$ are associated with $(q-k)^2$ and $(\vec{q}-\vec{k})^2$, 
respectively, we find that
\bea
J_{r,m,a,b}
&=&\frac{(-1)^{\sum_{i}\lambda_i+\sum_{j}\rho_j+1}}{\prod_{i=1}^{r}\prod_{j=r+1}^{r+m}\prod_{l=1}^a\prod_{u=a+1}^{a+b}\Gamma(\lambda_i)\Gamma(\rho_j)\Gamma(\alpha_l)\Gamma(\beta_u)}\nonumber\\
&&\int_0^1\prod_{i=0}^{r}dx_i\,x_i^{\lambda_i-1}\prod_{j=r+1}^{r+m}dx_j\,x_j^{\rho_j-1}\prod_{l=0}^{a}dy_l\,y_l^{\alpha_l-1}\prod_{u=a+1}^{a+b}dy_u\,y_u^{\beta_u-1}\nonumber\\
&&\delta(1-\sum_{j=0}^{r+m}x_j-\sum_{l=0}^{a+b}y_l)\int_0^{\infty}dz\,z^{\xi(r,m,a,b)+1}\cdot J_l[1]\;,\label{11}\\
&&\nonumber\\
\mbox{where}&&\nonumber\\
J_l[1]&=&\int\frac{d^{2n}l}{i^2\pi^{n}}\,\exp\Big\{-z\bigl[l_4{\cal M}_4l_4+\vec{l}{\cal M}\vec{l}-2{\cal N}_4l_4-2{\cal N}\vec{l}+{\cal R} \bigr] \Big\}\nonumber\\
&=&\frac{\exp\big\{-z{\cal R} \big\}}{\pi^{2(\sigma+\omega)}}\cdot\int d^{4\sigma}l_4\exp\Big\{-z\bigl[l_4{\cal M}_4l_4-2{\cal N}_4l_4\bigr] \Big\}\nonumber\\
&&\cdot \int d^{4\omega}\vec{l}\exp\Big\{-z\bigl[\vec{l}{\cal M}\vec{l}-2{\cal N}\vec{l} \bigr] \Big\}\;,\label{12}\\
&&\nonumber\\
l_4&=&\left(\begin{array}{l}q_4\\k_4\end{array}\right)\quad ,\quad \vec{l}=\left(\begin{array}{l}\vec{q}\\\vec{k}\end{array}\right)\;,\nonumber\\
{\cal M}_4&=&\left(\begin{array}{ll}\sum_{j=0}^r x_j&-x_0\\-x_0&x_0+\sum_{j=r+1}^{r+m}x_j\end{array}\right)\;,\nonumber\\
{\cal M}&=&\left(\begin{array}{ll}\sum_{j=0}^rx_j+\sum_{l=0}^ay_l&-x_0-y_0\\-x_0-y_0&x_0+y_0+\sum_{j=r+1}^{r+m}x_j+\sum_{l=a+1}^{a+b}y_l\end{array}\right)\;,\nonumber\\
{\cal N}_4&=&\left(\begin{array}{l}\sum_{j=1}^rx_j\,p_{4,j}\\\sum_{j=r+1}^{r+m}x_j\,p_{4,j}\end{array}\right)\nonumber\;,\\
{\cal N}&=&\left(\begin{array}{l}\sum_{j=1}^rx_j\,\vec{p_{j}}+\sum_{l=1}^ay_l\,\vec{s_l}\\\sum_{j=r+1}^{r+m}x_j\,\vec{p_{j}}+\sum_{l=a+1}^{a+b}y_l\,\vec{s_l}\end{array}\right)\nonumber\;,\\
{\cal R}&=&\sum_{j=1}^{r+m}x_j\,p_j^2+\sum_{l=1}^{a+b}y_l\,\vec{s_l}^2\;,\nonumber\\
&&\nonumber\\
\xi(r,m,a,b)&=&\sum_{i=1}^r\lambda_i+\sum_{j=r+1}^{r+m}\rho_j+\sum_{l=1}^{a}\alpha_l+\sum_{u=a+1}^{a+b}\beta_u\quad ,\quad \lambda_0=\alpha_0=1\;.\nonumber
\eea
Observe that  for the generic case 
$\lambda_i=\rho_j=\alpha_l=\beta_u=1$, $\xi(r,m,a,b)=r+m+a+b$. 
Integration over  $l$ in (\ref{12}) leads to
\bea
J_l[1]&=&z^{-2(\sigma+\omega)}\left(\mbox{Det}{\cal M}_4\right)^{-\frac{1}{2}}\left(\mbox{Det}{\cal M}\right)^{-\frac{1}{2}}\nonumber\\
&&\cdot\exp\Big\{-z\bigl[{\cal R}-{\cal N}_4{\cal M}_4^{-1}{\cal N}_4-{\cal N}{\cal M}^{-1}{\cal N}\bigr] \Big\}\;,\label{J1}\\\
\mbox{where}&&\nonumber\\
\mbox{Det}{\cal M}_4&=&\Bigl[x_0\cdot \sum_{j=1}^{r+m}x_j+(\sum_{j=1}^{r}x_j)(\sum_{j=r+1}^{r+m}x_j)  \Bigr]^{2\sigma}\;, \label{detm4}\\
\mbox{Det}{\cal M}&=&\Bigl[(x_0+y_0)(\sum_{j=1}^{r+m}x_j+\sum_{l=1}^{a+b}y_l)+(\sum_{j=1}^{r}x_j+\sum_{l=1}^{a}y_l)(\sum_{j=r+1}^{r+m}x_j+\sum_{l=a+1}^{a+b}y_l)  \Bigr]^{2\omega}\;. \nonumber
\eea

By differentiating the integral (\ref{12}) repeatedly with respect to 
${\cal N}_4$ and/or ${\cal N}$, we may easily derive the appropriate expression for
 any tensor integral.\\
Denoting  the  integrals with non-trivial numerators by
$J_{l}[l_4^{(i)}],J_{l}[\vec{l}^{(i)}],J_{l}[l_4^{(i)}l_4^{(j)}],\ldots$ etc.,
where $l_4^{(1)}=q_4,l_4^{(2)}=k_4,\vec{l}^{(1)}=\vec{q},\vec{l}^{(2)}=\vec{k}$, 
and applying  result (\ref{J1}), we get:
$$
\frac{\partial}{\partial {\cal N}_4^{(i)}}J_{l}[1]=2z\,l_4^{(i)}J_{l}[1]=2z\,J_{l}[l_4^{(i)}]
=2z\left({\cal N}_4{\cal M}_4^{-1}\right)^{(i)}J_{l}[1]\;.
$$
The procedure for other  integrals is similar, leading to expressions such as
\bea
J_{l}[l_4^{(i)}]&=&\left({\cal N}_4{\cal M}_4^{-1}\right)^{(i)}J_{l}[1]\;,\nonumber\\
J_{l}[\vec{l}^{(i)}]&=&\left({\cal N}{\cal M}^{-1}\right)^{(i)}J_{l}[1]\;,\nonumber\\
J_{l}[l_4^{(i)}\vec{l}^{(j)}]&=&\left({\cal N}_4{\cal M}_4^{-1}\right)^{(i)}\left({\cal N}{\cal M}^{-1}\right)^{(j)}J_{l}[1]\;,\nonumber\\
J_{l}[l_4^{(i)}l_4^{(j)}]&=&\Big\{\left({\cal N}_4{\cal M}_4^{-1}\right)^{(i)}\left({\cal N}_4{\cal M}_4^{-1}\right)^{(j)}+\frac{1}{2z}\left({\cal M}_4^{-1}\right)^{(ij)}\Big\}J_{l}[1]\;,\nonumber\\
J_{l}[\vec{l}^{(i)}\vec{l}^{(j)}]&=&\Big\{\left({\cal N}{\cal M}^{-1}\right)^{(i)}\left({\cal N}{\cal M}^{-1}\right)^{(j)}+\frac{1}{2z}\left({\cal M}^{-1}\right)^{(ij)}\Big\}J_{l}[1]\;,\nonumber\\
J_{l}[l_4^{(i)}l_4^{(j)}\vec{l}^{(k)}]&=&\Big\{\left({\cal N}_4{\cal M}_4^{-1}\right)^{(i)}\left({\cal N}_4{\cal M}_4^{-1}\right)^{(j)}\left({\cal N}{\cal M}^{-1}\right)^{(k)}\nonumber\\
&&+\frac{1}{2z}\left({\cal M}_4^{-1}\right)^{(ij)}\left({\cal N}{\cal M}^{-1}\right)^{(k)}\Big\}J_{l}[1]\;,\label{nums}\\
&&\vdots\nonumber
\eea
To cope with integrals in which either the factor $(q-k)^2$, or 
$(\vec{q}-\vec{k})^2$, is absent in the basic integral (\ref{Itwo}), we proceed by 
first introducing the indicator functions
\be
i_B=\Bigl\{\begin{array}{ll}1&\mbox{ if } (q-k)^2 \mbox{ is present }\\0&\mbox{ if } (q-k)^2 \mbox{ is not present }\end{array}\; ,\;
i_C=\Bigl\{\begin{array}{ll}1&\mbox{ if } (\vec{q}-\vec{k})^2 \mbox{ is present }\\0&\mbox{ if } (\vec{q}-\vec{k})^2 \mbox{ is not present }\end{array}\label{ib}
\ee
and 
$$
\mathrm{ind}(i_B)=\Bigl\{\begin{array}{ll}\delta(x_0)&\mbox{ for } i_B=0\\1&\mbox{ for } i_B=1\end{array}\quad ,\quad
\mathrm{ind}(i_C)=\Bigl\{\begin{array}{ll}\delta(y_0)&\mbox{ for } i_C=0\\1&\mbox{ for } i_C=1\end{array}\;,
$$
such that  $x_0$ is set to zero if  $(q-k)^2$ is absent, and  $y_0$ is set 
to zero if $(\vec{q}-\vec{k})^2$ is absent. 
After integration over $z$, 
the most general formula for the scalar (basic) integral (\ref{11}) is then given by
\bea
J_{r,m,a,b}
&=&(-1)^{\sum_{i}\lambda_i+\sum_{j}\rho_j+i_B}\frac{\Gamma(\xi(r,m,a,b)+i_B+i_C-2\sigma-2\omega)}{\prod_{i=1}^{r}\prod_{j=r+1}^{r+m}\prod_{l=1}^a\prod_{u=a+1}^{a+b}\Gamma(\lambda_i)\Gamma(\rho_j)\Gamma(\alpha_l)\Gamma(\beta_u)}\nonumber\\
&&\int_0^1\prod_{i=0}^{r}dx_i\,x_i^{\lambda_i-1}\prod_{j=r+1}^{r+m}dx_j\,x_j^{\rho_j-1}\prod_{l=0}^{a}dy_l\,y_l^{\alpha_l-1}\prod_{u=a+1}^{a+b}dy_u\,y_u^{\beta_u-1}\nonumber\\
&&\delta(1-\sum_{j=0}^{r+m}x_j-\sum_{l=0}^{a+b}y_l)\cdot \mathrm{ind}(i_B)\cdot \mathrm{ind}(i_C) 
\cdot \left(\mbox{Det}{\cal M}_4\right)^{-\frac{1}{2}}\left(\mbox{Det}{\cal M}\right)^{-\frac{1}{2}}\nonumber\\
&&\Bigl[{\cal R}-{\cal N}_4{\cal M}_4^{-1}{\cal N}_4-{\cal N}{\cal M}^{-1}{\cal N}\Bigr]^{2\sigma+2\omega-\xi(r,m,a,b)-i_B-i_C} \;.
\label{Iscal}
\eea
The result in Minkowski space is obtained 
by making the replacements $ip_4 \to p_0$ and $p_E^2 \to -p^2$.\\
Although the expressions given above look somewhat complicated, 
since they are intended to represent all types of integrals that may occur in a 
two-loop calculation, they can be easily implemented into 
an algebraic manipulation 
program in order to yield the parameter representations of the various integrals.

\end{document}